\documentclass[12pt]{iopart}

\usepackage[numbers,sort&compress]{natbib}
\usepackage{lineno}
\usepackage{graphicx}
\usepackage[labelfont=bf,font=scriptsize]{caption}
\usepackage{xcolor,colortbl}
\usepackage{ulem}
\usepackage{hyperref}
\usepackage{booktabs} 
\usepackage{changepage,threeparttable} 

\definecolor{magenta}{rgb}{1.0,0.0,1.0}
\definecolor{green}{rgb}{0.0,0.5,0.0}
\definecolor{gray}{gray}{0.9}

\usepackage{iopams}
\usepackage{siunitx}
\makeatletter
\long\def\@makefntext#1{\parindent 1em\noindent 
 \makebox[1em][l]{\footnotesize\rm$\m@th{^\arabic{footnote}}$}%
 \footnotesize\rm #1}
\def\@makefnmark{\hbox{${^\arabic{footnote}}\m@th$}}
\def\@thefnmark{\arabic{footnote}}
\makeatother
\begin{document}
\title{Machine-learning enhanced dark soliton detection in Bose-Einstein condensates}

\author{Shangjie Guo\textsuperscript{1}, Amilson R. Fritsch\textsuperscript{1}, Craig Greenberg\textsuperscript{2}, I.~B.~Spielman\textsuperscript{1}, Justyna P. Zwolak\textsuperscript{2,*}}
\address{\textsuperscript{1}Joint Quantum Institute, National Institute of Standards and Technology, and University of Maryland, Gaithersburg, MD 20899, USA}
\address{\textsuperscript{2}National Institute of Standards and Technology, Gaithersburg, MD 20899, USA}
\ead{jpzwolak@nist.gov}

\begin{abstract}
Most data in cold-atom experiments comes from images, the analysis of which is limited by our preconceptions of the patterns that could be present in the data.
We focus on the well-defined case of detecting dark solitons---appearing as local density depletions in a Bose-Einstein condensate (BEC)---using a methodology that is extensible to the general task of pattern recognition in images of cold atoms. 
Studying soliton dynamics over a wide range of parameters requires the analysis of large datasets, making the existing human-inspection-based methodology a significant bottleneck. 
Here we describe an automated classification and positioning system for identifying localized excitations in atomic BECs utilizing deep convolutional neural networks to eliminate the need for human image examination.
Furthermore, we openly publish our labeled dataset of dark solitons, the first of its kind, for further machine learning research.
\end{abstract}
\vspace{2pc}

\section{Introduction}\label{sec:intro}
Machine-learning (ML)-based image classification has found application throughout science, from analysis of experimental data in particle physics~\cite{Aurisano_2016,Acciarri_2017,kagan2020imagebased}, dark matter search experiments~\cite{Golovatiuk_2020,Khosa_2020} or quantum dots experiments~\cite{Kalantre17-MLD,Mills19-CAT,Zwolak20-AQD,Usman2020} to predicting properties of materials~\cite{cao2019convolutional,PhysRevMaterials.4.093801,gubernatis2018machine} to studying molecular representations and properties~\cite{duvenaud2015convolutional,butler2018machine,winter2019learning}. 
In atomic physics, ML has been used to locate topological phase transitions~\cite{Rem19-PTN}, to complement absorption imaging technique~\cite{PhysRevApplied.14.014011}, to characterize particles in disordered fields~\cite{Pilati19-SMU}, and to detect quantum vortices in rotating BECs~\cite{metz2021deep}. In this paper, by combining convolutional neural networks (ConvNets) with traditional fitting techniques, we first categorize many-body atomic physics data, and then extract quantitative information from this data.

Using cold-atom Bose-Einstein condensates (BECs), we focus on solitons, robust solitary waves that retain their size, shape, and speed at which they travel~\cite{PhysRevLett.101.130401,Frantzeskakis_2010}.
These properties arise from an interplay between nonlinearity and dispersion that is present in many physical systems. 
Indeed, since their first observation in canals~\cite{Russel1837}, solitons have been found in rivers and seas~\cite{Osborne1980, Lakshmanan2009}; BECs~\cite{Burger1999,Denschlag97}; optical fibers~\cite{Hasegawa1973, PhysRevLett.45.1095}; astronomical plasmas~\cite{Stasiewicz2003}; and even human blood vesicles~\cite{Hashizume1985,Yomosa1984}. 
Due to their inherent stability, solitons in optical fibers~\cite{MOLLENAUER20061} have found commercial applications in long-distance, high-speed transmission lines~\cite{902164}.

While the natural environment does not allow for the controlled study of quantum solitons, BECs are an excellent medium where individual or multiple solitons can be created on-demand, with all their properties, such as position and velocity, tuned according to necessity~\cite{Aycock2503, Fritsch20-SCV}. 
Most measurements in BEC experiments produce raw data in the form of images that, in our context, provide information about the solitons' positions within the BEC. 
The challenge is to efficiently and reliably identify the number of solitons and their locations. 
Traditional least-squares fitting techniques can locate solitons, provided that the soliton number is known in advance.
Currently, the number of solitons is determined manually~\cite{Fritsch20-SCV}, and this human intervention inhibits the automated analysis of large datasets.

\begin{figure}
    \flushright
    \includegraphics[width=0.99\linewidth]{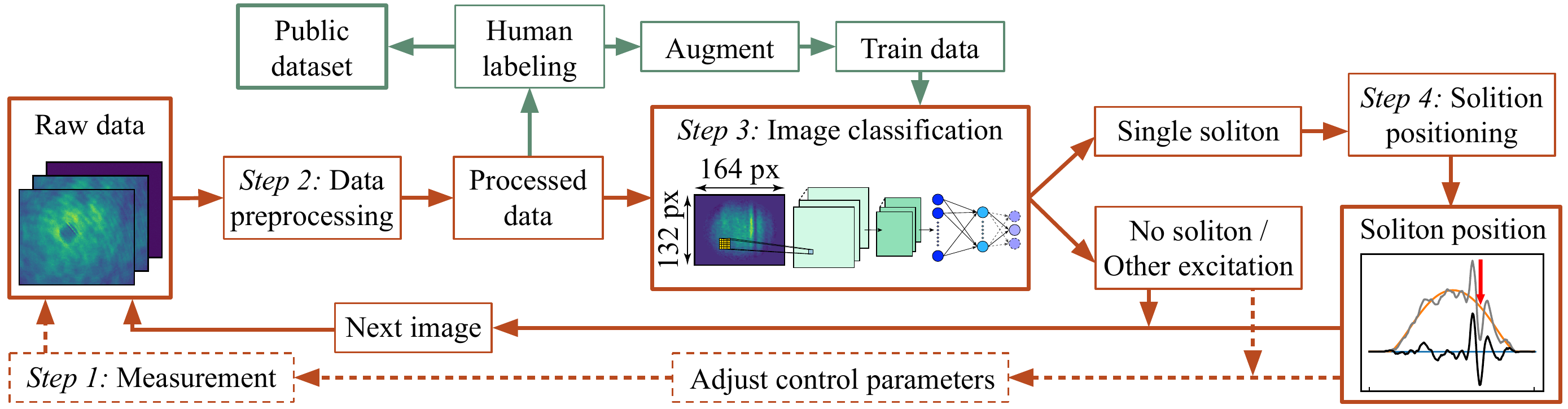}
    \caption{Schematic of the soliton detection and positioning system. 
    Red boxes and arrows represent the flow of the full system. 
    The dashed red boxes and arrows represent additional components required for a closed-loop implementation. 
    The green boxes and arrows represent additional out-of-loop steps of preparing the classifier and establishing the training dataset.
    }
    \label{fig:pipeline}
\end{figure}

Here, we describe our reliable automated soliton detection and positioning system that takes as input image data and outputs information whether a single soliton is present, and, if so, its location.
Since solitons are easily identifiable by human examination of images, this problem naturally connects to the field of computer vision and ConvNet-based image classification~\cite{rawat2017deep}. 
Our algorithm consists of a data preprocessor that converts raw data into a ConvNet-compatible format; a ConvNet image classifier that determines if a single soliton has been detected; and a position regressor that locates the soliton within the BEC, when applicable (see figure~\ref{fig:pipeline} for a schematic of the analysis flow).

We show that our fully automated system performs comparably to our existing human image classifier, autonomously replicating the data analysis in Ref.~\cite{Fritsch20-SCV}.
In addition to developing a detection and positioning tool, we established a dataset of over 6\,000 labeled experimental images of BECs with and without solitonic excitations; this dataset is available via the National Institute of Standards and Technology (NIST) Science Data Portal~\cite{solitons-data} and at data.gov.

The remainder of this paper is organized as follows: in section~\ref{sec:ml}, we illustrate the work flow of the soliton detector and its preparation process.
Then in section~\ref{sec:result}, we demonstrate the system, quantify its performance, and discuss the quality of the labeled dataset.
Finally in section~\ref{sec:conclusion}, we conclude and discuss possible future directions.

\section{Soliton detection and position system}\label{sec:ml}
In this section we describe our fully automated method of soliton detection and positioning in images of BECs.
Our four-step protocol, detailed in the following sections and depicted in figure~\ref{fig:pipeline}, is outlined as follows.

\noindent{\it Step 1: Measurement.} 
The measurement consists of three raw images that are combined to produce a single image of the atomic density distribution.

\noindent{\it Step 2: Data preprocessing.}  
As shown in figure~\ref{fig:pipeline}, the BEC is rotated with respect to the image frame orientation, and the region of interest where atoms are captured is a small fraction of the full image.
To simplify soliton positioning, the data is first rotated to align the BEC orientation with the image frame and then cropped prior to the classification step.

\noindent{\it Step 3: Image classification.} 
The pre-trained ConvNet classifier determines whether a lone soliton is present in a given image. If so, step four is executed, otherwise the image analysis terminates.

\noindent{\it Step 4: Soliton positioning.} 
The soliton position with respect to the BEC center is determined using a least-squares fit based on a one-dimensional (1D) model function.

\subsection{Experimental setup and measurement}\label{ssec:exp}
In our experiments, solitons are created and propagate the nonlinear media of a $^{87}\textrm{Rb}$ atomic BEC. We create BECs using well-established techniques for cooling and trapping atoms~\cite{Lin2009}, allowing us to obtain $N=2.4(2)\times10^5$ atom\footnote[3]{We use a notation value(uncertainty) to express uncertainties, for example $1.5(6)\ {\rm cm}$ would be interpreted as $(1.5\pm0.6)~{\rm cm}$. All uncertainties herein reflect the uncorrelated combination of single-sigma statistical and systematic uncertainties.} condensates in a time-averaged crossed optical dipole trap.
Since solitons are only stable in quasi-1D systems~\cite{PhysRevLett.89.110401}, i.e., resulting from highly anisotropic trapping geometries, our potential is elongated, with trapping frequencies $[\omega_x,\omega_y,\omega_z] = 2\pi\times[9.1(1),153(1),94.5(6)]\ \si{\hertz}$.

We launch solitons using our recently developed `improved' protocol, that simultaneously engineers the density and phase of the BEC wave function~\cite{Fritsch20-SCV}. 
By contrast with the `standard' protocol that only modifies the BEC phase and can only create solitons within a small range of initial velocities, our protocol can create solitons with arbitrary initial velocity. 
The potentials for density engineering and phase imprinting are both generated by far-detuned laser light, spatially patterned by a digital micromirror device (DMD). Our protocol is summarized as follows: After the BEC is created, we reduce its local density by applying a repulsive dimple potential. Next, the DMD is reprogrammed to display a step function that illuminates only half of the BEC, imprinting the soliton's phase profile.
To minimize creating additional density perturbations, the dimple potential is reapplied and its magnitude slowly ramped to zero.
We note that in our data there are additional solitonic excitations that, while representing different physical states (e.g., kink solitons, solitonic vortices, soliton rings and so forth~\cite{Mateo2015}), can result in similar image and we identify simply as solitons in our analysis.

After solitons are created, we let them oscillate in the harmonic trapping potential for a variable evolution time.
For evolution times much less than the trap period, additional density excitations from the soliton imprinting process are present.
We then turn off the trapping potential and let the BEC evolve for a $15\ \si{\milli\second}$ time of flight, before absorption imaging the resulting density distribution~\cite{ketterle1999making}.

\subsection{Data preprocessing}\label{ssec:prep}
We established a dataset of over $6.2\times10^3$ images for ConvNet training; these images were taken from multiple experiments performed in a single lab over a span of two months.
The raw images were obtained with a 648$\times$488 pixel camera (Point Grey FL3) with $5.6\ \si{\micro\meter}$ square pixels, labeled by $i$ and $j$. Including the $\approx 6\times$ magnification, each pixel has effective $0.93\ \si{\micro\meter}$ size. The diffraction limit of the imaging system gives an optical resolution of $\approx 2.8\ \si{\micro\meter}$ (roughly three pixels).  

Absorption imaging combines three raw images into a single record of atomic density.
In the first image $I^{\rm A}_{i,j}$, a probe laser illuminates the BEC and the resulting intensity records the probe with the BEC's shadow.
The second image $I^{\rm P}_{i,j}$ records only the probe intensity, and the third image $I^{\rm BG}_{i,j}$ is a dark frame containing any ambient background signal.
The 2D column density
\begin{equation}\label{eq:od}
    \sigma_0 n_{i,j} \approx -\ln\left[\frac{I^{\rm A}_{i,j}-I^{\rm BG}_{i,j}}{I^{\rm P}_{i,j}-I^{\rm BG}_{i,j}}\right]
\end{equation}
can be derived from these images, where the resonant cross-section $\sigma_0=3\lambda^2 / (2\pi)$ is derived from the wavelength $\lambda$ of the probe laser.
The dimensionless product $\sigma_0 n_{i,j}$ is of order $1$ in our data, so we express density in terms of this product.
Figure~\ref{fig:pipeline} shows an example of the probe beam with atoms and the resulting density in the `raw data' and `image classifier' frames, respectively. 

In our raw data, the BEC occupies only a small region of the image, and the long axis of the BEC is rotated by about 43 degrees with respect to the camera. 
To facilitate the ConvNet training, the images are rotated to align the BEC with the image frame and cropped to discard the large fraction of the image that does not contain information about the BEC. 
Since the BEC's position and shape can vary for different realizations of the same experiment, we implement a fitting approach to determine the position and size of the BEC.

Next, we fit every image to a column-integrated 3D Thomas-Fermi distribution~\cite{Castin1996}, giving the 2D distribution:
\begin{equation}\label{eq:2Dfit}
n^{\rm TF}_{i,j} = n_{0}\,{\max}\left\{\left[1-\left(\frac{i-i_0}{R_i}\right)^2-\left(\frac{j-j_0}{R_j}\right)^2\right],0\right\}^{3/2} + \delta n.
\end{equation}
This function describes the density distribution of 3D BECs integrated along the imaging axis.
We use six parameters to fit: the BEC center coordinates $[i_0, j_0]$; the peak 2D density $n_0$; the Thomas-Fermi radii [$R_i$, $R_j$]; and an offset $\delta n$ from small changes in probe intensity between images.

Successful fitting requires acceptable initial guesses for all fit parameters.
We obtained guesses for $i_0$ and $j_0$ by summing the image along the vertical and horizontal directions to obtain two 1D projections, from which we select the average position of the five largest values as the initial guesses.
We took the largest value of the image as the guess for $n_0$ and used $[R_i, R_j] = [66,55]$ pixels, based on the typical radii over the whole dataset.
The guess for the offset $\delta n$ is zero.
The result of these fits are included in our released dataset.

\begin{figure}[t]
    \flushright
    \includegraphics[width=0.99\linewidth]{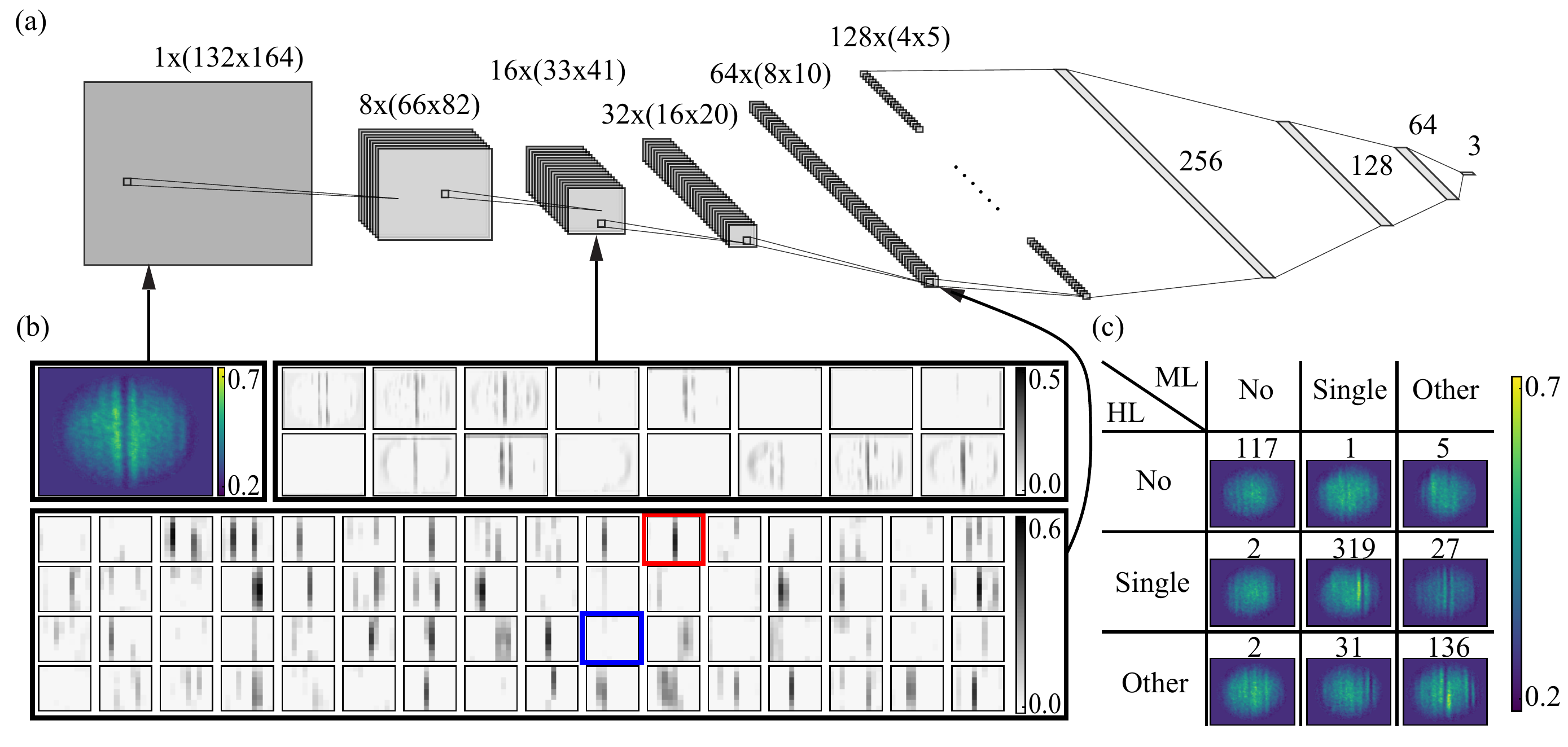}
    \caption{
    (a) ConvNet classifier structure. 
    The first box represents a preprocessed input image. 
    Each of the following left-most five architecture components represent a combination of a convolutional layer with a ReLU activation function and a max pooling layer, with their filter number and image size specified.
    Each of the following three components represent a combination of a fully connected layer with a ReLU activation, and a dropout layer, with their neuron number specified. 
    The last component represents a fully connected output layer with softmax activation. 
    (b) Visualization of the input, second, and fourth max pooling layer activation for a successfully classified single soliton image. 
    The top left panel is the input image, the 16 images in the top right panel are the output of the max pooling layer, and 64 images in the bottom panel are the output of the fourth max pooling layer. 
    The red boxed filter indicates one of the filters that captures the lone soliton feature. 
    The blue boxed filter would activate if more than one soliton is present (see appendix~\ref{ssec:02visual} for no soliton/other excitation).  
    (c) Confusion matrix of the test set, comparing between human assigned labels (HL) and ML classifier prediction (ML). 
    The images show sample successful (diagonal) and misclassified (off-diagonal) cases. 
    The numbers above indicate how many images are assigned to a given class.
    }
    \label{fig:soliton-sample}
\end{figure}

We determined the $164\times132$ pixel extent of the cropping region by examining the radii $[R_i, R_j] = [66(5), 58(3)]$ obtained from fits to $6.2\times10^3$ images.
We then centered the cropping region at $[i_0, j_0]$ as determined from fits of each image separately.
The process was validated on an additional $10^4$ images not included in our dataset.
In the preprocessed images, dark solitons appear as vertically aligned density depletions and are easily visually identified (see top-left panel in figure\ref{fig:soliton-sample}(b)).

\subsection{Labeling}\label{ssec:labeling}
Three independent human labelers labeled the preprocessed data, categorizing the images into three classes: `no soliton', `single soliton', and `other excitations'.
The `no soliton' class contains images that unambiguously contains no solitons;
the `single soliton' class describes images with one and only one soliton;
and `other excitations' class covers any image that can neither be interpreted as `no soliton' nor `single soliton.'
We did not include a separate `two soliton' class in our demonstration because the small number of images with two solitons led to ineffective training.

The labeling process was carried out in eight batches, with each batch size limited by the attention span of the labelers.
Once a given batch was completed, the resulting labels were compared and images with full agreement were set aside. 
The overall labeling agreement rate was $87\,\%$ (table~\ref{tab:1} shows a comparison of the labeling agreement for all three classes), consistent across all batches. 
The remaining images were further analyzed and discussed until an agreement was reached. 
The final distribution of images between classes is as follows: $19.8\,\%$ in the no soliton class, $55.4\,\%$ in the single soliton class, and $24.8\,\%$ in the other excitations class. 
Figure~\ref{fig:soliton-sample}(c) shows representative labeled images from each class. 
This labeled dataset was employed to train the ConvNet classifier and to test the positioning protocol. 

\begin{table}[t]\flushright
\scriptsize
\captionsetup{width=\linewidth,font=footnotesize}
\caption{Human labeling result. The first two columns (Full) show image counts and percentages of each class. The last two columns (3-agree) compare the counts and ratio in the all data of each class for the images with labels that humans initially agreed on.}
\begin{tabular*}{\textwidth}{@{} l@{\extracolsep{\fill}}cccccc@{}}\toprule
Dataset & &\multicolumn{2}{c}{Full} & &\multicolumn{2}{c}{3-agree} \\
Class & & Count & Percentage [\%] & & Count &  Agreement ratio [\%] 
\\ \midrule
No soliton & & 1\,237 & 19.8 & & 1\,184 & 95.7  
\\
Single soliton & & 3\,468 & 55.4 & & 3\,077 & 88.7 
\\
Other excitations & & 1\,552 & 24.8 & & 1\,184 & 76.3  
\\ \midrule
Total & & 6\,257 & 100.0 & & 5\,445 & 87.0 
\\
\bottomrule
\end{tabular*}
\label{tab:1}
\end{table}

\subsection{Image classification}\label{ssec:class}
Our ConvNet classifier, shown in figure~\ref{fig:soliton-sample}(a), consists of five convolutional layers.
Each layer is followed by a rectified linear unit (ReLU) function defined as $f(x)=\max(0,x)$, then a max pooling layer\footnote[4]{Max pooling is a form of non-linear down-sampling that converts the input ($km\times ln$) array, partitioned into a set of non-overlapping rectangles of equal ($k\times l$) size, into a smaller ($m\times n$) array with entries representing the maximum value of the corresponding sub-region.}. 
The final max pooling layer is flattened and fully connected to a deep neural network with three hidden layers (256, 128, and 64 neurons, respectively) and an output layer (three neurons). 
Each hidden layer is followed by the ReLU activation function, and to reduce overfitting, a dropout layer that randomly eliminates neural connections with a frequency of $0.5$ during each training stage.
The output vector ${\boldsymbol \xi} = (\xi_1, \xi_2, \xi_3)$ is normalized by the softmax activation function, giving the final output probabilities $P_m({\boldsymbol \xi})=\exp(\xi_m)/\sum_n \exp(\xi_n)$.

\begin{table}[t]\centering
\scriptsize
\captionsetup{width=\linewidth}
\caption{Classification performance summary for the best classifier when training with the full training dataset with performance measured using cross-validation from the training, when testing on the full test dataset, and when testing on a subset of the test dataset with labels that labelers initially agreed on.}
\begin{tabular*}{\textwidth}{l@{\extracolsep{\fill}}cccc}\toprule
& & Cross-validation & Full test set & Labelers initially agreed subset\\\midrule
Accuracy [\%] & &89.6(5) &89.4 &91.6 \\
Weighted F1 & &0.896(6) &0.894 &0.916 \\
No soliton F1 & &0.938(10) &0.959 &0.983 \\
Single soliton F1 & &0.920(4) &0.913 &0.935 \\
Other excitations F1 & &0.806(6) &0.807 &0.782 \\
\bottomrule
\end{tabular*}
\label{tab:2}
\end{table}

The labeled dataset was divided into two subsets: 640 images ($10.2\,\%$ of the dataset) were set aside as testing set, while the remaining 5\,617 images ($89.8\,\%$) were used for training during the model architecture development. 
Since our training dataset is unbalanced, i.e., its different classes have a significantly different number of images, we balance it using augmentation techniques.
We augment using three physically acceptable transformations: horizontal and vertical reflections, as well as a 180 degree rotation. 
All three transformation were applied to the no soliton and other excitations classes, increasing their size by a factor of four. 
For the single soliton class we used one randomly chosen transformation per image, doubling the size of this class. 
After augmentations, the size of the three classes has a $0.28: 0.38 :0.34$ fractional distribution.
To model a small rotation angle present in different realizations of our BEC, we randomly rotate images by an angle in the range $\pm 1$ degree every time they are used during the training process. 
We applied an elliptical mask with radii $[R_i, R_j]$ to each image, eliminating all technical noise outside the BEC, to accelerate the training process\footnote[5]{In early training attempts the classifier learned to separate the BEC from the background. Because the BEC resides within a well-defined ellipse we accelerated convergence by applied the elliptical mask prior to training.}.
Lastly, we preconditioned the data to have a range suitable for ConvNet input by uniformly scaling the image-values to the $[0,1]$ range.

Since our testing dataset remains unbalanced, we assess the performance of trained models using the weighted F1 score~\cite{scikit-learn}. 
When two models have similar weighted F1 scores, we first compare their accuracies as a tie-breaker, and if that fails we use the F1 score of the single soliton class\footnote[6]{We use the F1 score of the single soliton class as the final tie-breaker, because we ultimately compare to single-soliton dynamics data in section~\ref{ssec:detector}.}.

We used a semi-structured search through the model parameter space, and the resulting performance for varying hyperparameters is detailed in the appendix~\ref{ssec:tuning}.
Once we determined the best performing model, we used randomly selected $95\,\%$ of training set for the final training.
Training terminated when the F1 score of the remaining $5\,\%$ did not increase for five epochs.
We took the model prior to these five non-improving epochs as our final trained model.

Figure~\ref{fig:soliton-sample}(b) shows representative intermediate convolutional layers of the trained model, with a correctly classified single soliton as the input. 
We observe that some filters, such as the one marked with a red box, successfully capture the information of a single soliton (further examples are presented in appendix~\ref{ssec:02visual}).

Figure~\ref{fig:soliton-sample}(c) and the second column of table~\ref{tab:2} show the results of our final soliton classifier.
In summary, our model has weighted ${\rm F}_1\approx 0.9$ and accuracy $\approx 90\,\%$, in excess of the $87.0\,\%$ human agreement ratio. 
The most frequent classifier errors conflate images from the single soliton class and the other excitations class: $6.9\ \%$ of the single soliton images is wrongly assigned to the other classes ($P_1<0.2$), and $4.3\ \%$ has no clear assignment ($0.2\leq P_1<0.8$).

\begin{figure}
    \flushright
    \includegraphics[width=0.99\linewidth]{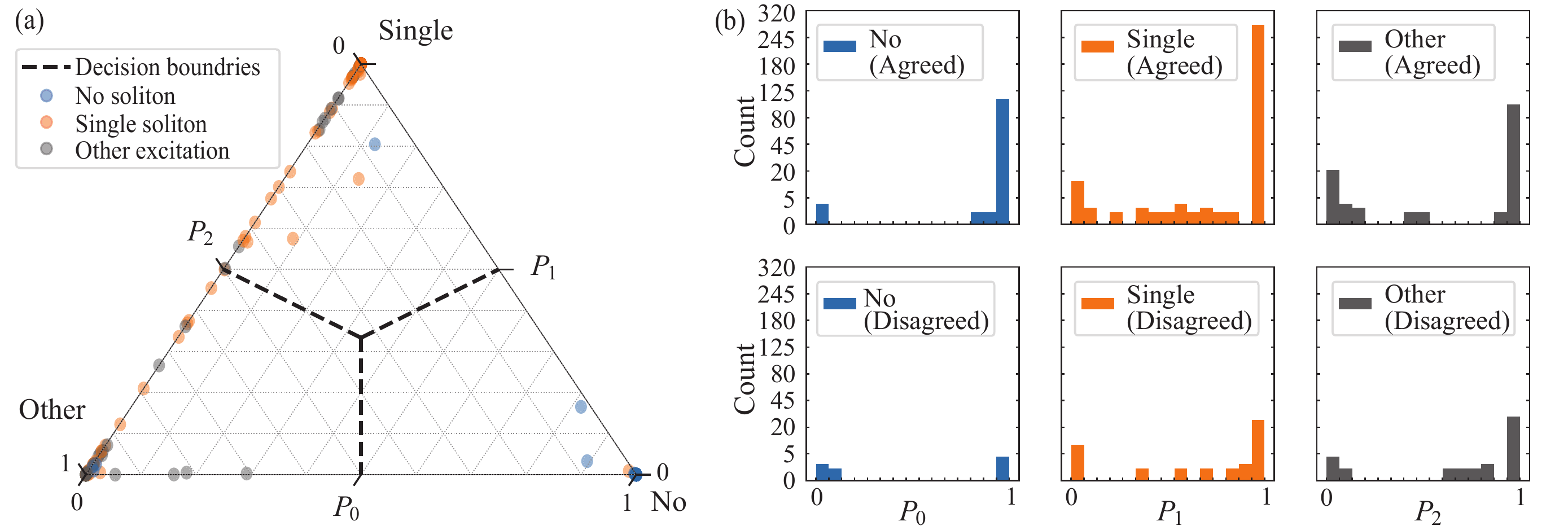}
    \caption{Soliton classification results. 
    (a) Distribution of test data, colored by ground truth label. 
    The scattered dots with different labels overlap each other in a randomized order. 
    (b) Histogrammed probabilities. 
    The upper panels histogram the classification probabilities from human-initially-agreed data, while the lower panels histogram those from human-initially-disagreed data. 
    The vertical axes are in square root scale to emphasize the misclassified data.
    }
    \label{fig:class-result}
\end{figure}

Figure~\ref{fig:class-result}(b) shows that the classifier works very well for the no soliton and single soliton classes.
The classifier performs better when tested against human-initially-agreed data than human-initially-disagreed data, suggesting that some disagreed upon images may be truly ambiguous (Also see the last column in table~\ref{tab:2}).
In addition, we observe an anomalously large misclassification rate for human agreed data in the other excitations class, resulting from the human labelers use of this class when facing a dilemma. 
Furthermore, the wrongly classified data are distributed near the corners of figure~\ref{fig:class-result}(a), indicating a high degree of confidence in the misclassification.

\subsection{Position regression}\label{ssec:position}
Once images containing only one soliton are identified, we locate the soliton position using a simple yet robust least-squares fitting procedure~\cite{scikit-learn}. 
The first step consists of summing each 2D image along the $j$ direction to obtain a 1D distribution $n_i = \sum_j n_{i,j}$.
We fit the 1D distributions to the expected profile:
\begin{eqnarray}
  n_{i}^{\rm 1D}  &= \, n_0^{\rm 1D} {\max}\left\{\left[1-\left(\frac{i-i_0}{R_i}\right)^2\right],0\right\}^{2} + \delta n^{\rm 1D},
\end{eqnarray}
that is, Eq.~(\ref{eq:2Dfit}) integrated along the $j$ direction.
The initial guess for $n_0^{\rm 1D}$ was the max of the integrated distribution, and the remaining guesses were taken from the successful 2D fit.
We subtract the fit from the integrated 1D profile to obtain the residuals $\Delta_i = n_i - n_{i}^{\rm 1D}$. 
Ideally, this procedure would result in a flat background containing a single dip, associated with the soliton, which we identified using a Gaussian fit\footnote[7]{We found that the Gaussian gave the highest quality of fit as compared to other peaked functions.}.
We use the minimum of $\Delta_i$ as the initial guess for the Gaussian amplitude, the minimum position as the initial center, $3$ pixels for the width, and zero for the offset.
This fit yielded the soliton width, amplitude and position.

\section{Results}\label{sec:result}
\subsection{Soliton detector}\label{ssec:detector}
\begin{figure}
    \flushright
    \includegraphics{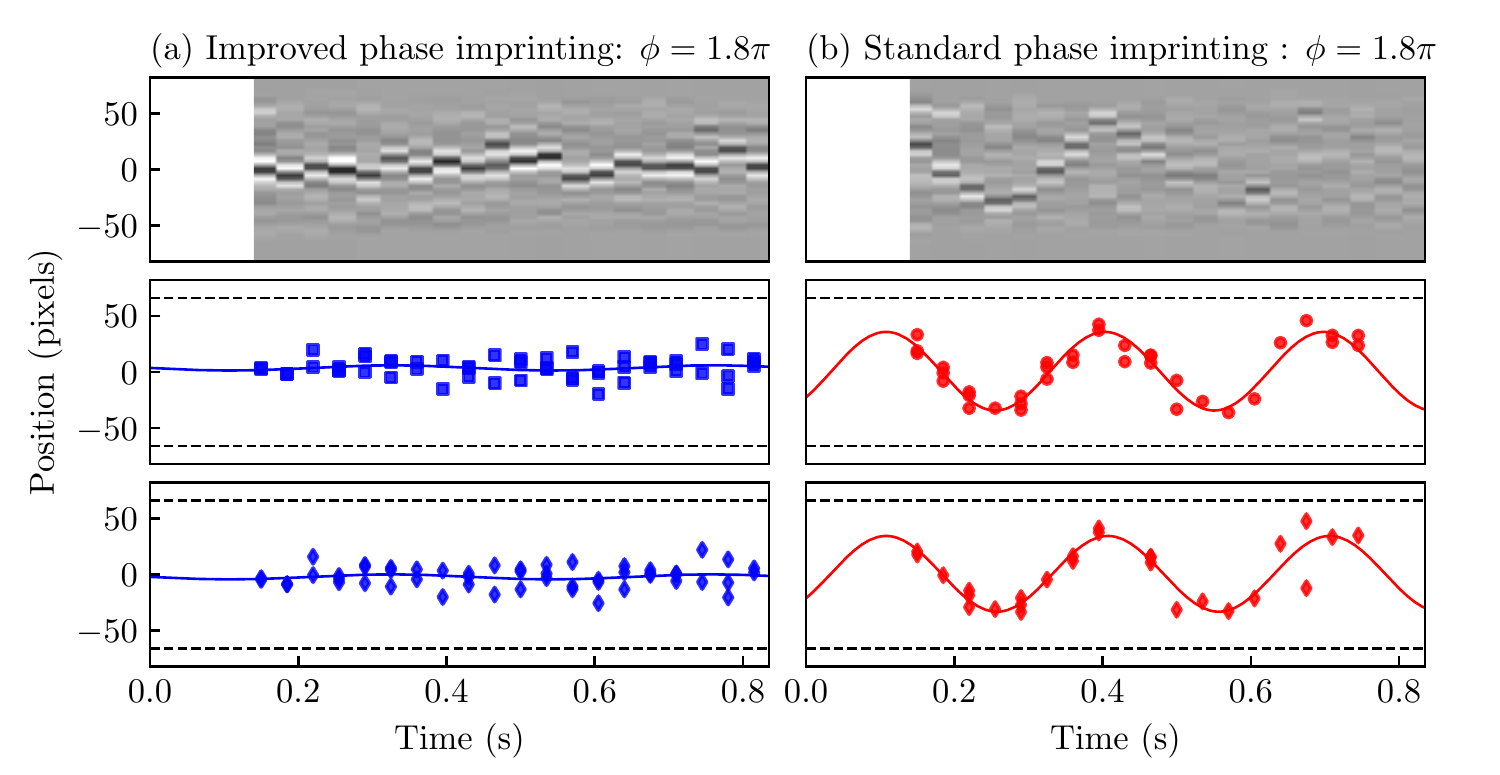}
    \caption{Oscillation of dark solitons created by applying $1.8(1)\pi$ phase using the (a) improved and (b) standard protocol described in~\cite{Fritsch20-SCV}. 
    Top panels show samples for the residuals $\Delta_i$, obtained after subtracting the fit from the 1D profile.
    Middle and bottom panels show the soliton positions and sinusoidal fits (as described in the text) based on manually identified images and the outputs of the automated system, respectively. 
    Dashed lines at $j=\pm66$ pixels in all four panels represent the edges of the BEC.}
    \label{fig:series_pos}
\end{figure}

To test the performance of the fully automated soliton detection and positioning system, we use two sets of images containing oscillating dark solitons\footnote[8]{These two sets of images contribute to the data published in figure 2 of Ref.~\cite{Fritsch20-SCV}, and are presented here in figure~\ref{fig:series_pos}.} that were launched using the standard and improved protocols described in section~\ref{ssec:exp}, with 60 and 59 images, respectively.

In the first test, we used the improved-protocol data-set, with representative summed data $n_i$ presented in the top panel of figure~\ref{fig:series_pos}(a).
As the solitons in these images are well pronounced, we expected the ConvNet will easily classify them.
Out of 59 images, 52 were classified as single soliton and the remaining 7 were classified as other excitations, in agreement with a human labeler. 
Solitons were then located in the first group by the positioning regressor (see figure~\ref{fig:pipeline}). 
The middle and bottom rows in figure~\ref{fig:series_pos}(a) plot the soliton position from manual and ConvNet identification, respectively.
We fit $i(t) = A \sin(\omega t + \Phi) + i_0$ to the soliton position data, and we compare the fitted parameters with those obtained from our previous manual approach.
As can be seen by comparing the middle and bottom rows of figure~\ref{fig:series_pos}(a), the performance of the automated protocol is basically indistinguishable from the manual approach.
The physical parameters from the ML classifier ($A = 2(2)$ pixels and $\omega/2\pi = 2.3(7)$ Hz) were within one standard deviation of those obtained for manual soliton identification ($A = 2(2)$ pixels and $\omega/2\pi = 2.3(6)$ Hz).

In the second test, we used images with solitons generated by the standard phase imprinting protocol.
As can be seen in top panel of figure~\ref{fig:series_pos}(b), solitons in these images can be shallower than those in figure~\ref{fig:series_pos}(a), making them potentially more difficult to distinguish from the no soliton and other excitations classes. 
Out of the 60 images in this test, 22 were classified by the ConvNet as no soliton, and 11 as other excitations, in agreement with a human labeler.
The remaining 27 were classified as a single soliton and were sent to the position regressor.
The lower panels in figure~\ref{fig:series_pos}(b) shows soliton position as a function of evolution time, obtained from manual~\cite{Fritsch20-SCV} and ConvNet identification, respectively.
Since \cite{Fritsch20-SCV} compared the soliton oscillation amplitude resulting from the two imprinting protocols, the authors did not limit themselves to images with a single soliton. 
Rather, when more than one soliton was created, the authors identified all the solitons but tracked only that associated with a specific trajectory. 
Since the ConvNet classifier was trained to select images with single soliton excitations, the middle panel in figure~\ref{fig:series_pos}(b) includes 12 more points than the bottom panel.
Even with fewer data points, however, the parameters from the ML classifier ($A = 34(3)$ pixels and $\omega/2\pi = 3.34(9)$ Hz) were within one standard deviation of those obtained for manual soliton identification  ($A = 35(2)$ pixels and $\omega/2\pi = 3.39(5)$ Hz).

The complete analysis resulting in both oscillation plots took under 148~\si{\second} per series on a 2014 MacBook Pro.
The expected performance relevant for in-situ operation, is $\approx2.4$~\si{\second} per image, a relatively small overhead on top of the measurement time (about 12~\si{\second}).  
In many cases, however, the analysis of an image would take place during the acquisition of the next image.

\subsection{Soliton dataset}\label{ssec:data}
As with all ML techniques, the availability of the training data is essential for good performance of the trained classifier. 
To assure the reliability of the assigned labels, the full dataset was independently labeled by three labelers, as described in section~\ref{ssec:prep}. 
Our full soliton image dataset consists of 6\,257 labeled images. There are 1\,237, 3\,468, and 1\,552 images for no soliton, single soliton, and other excitations classes, respectively. 

While for 5\,445 ($87.0\,\%$) of the images the assigned labels were consistent between labelers, for the remaining 812 images ($13.0\,\%$) there was a disagreement with at least one labeler. 
These images needed to be further discussed until an agreement was reached. As can be seen in table~\ref{tab:1}, the most challenging was distinguishing between images with single soliton and other excitations. This is likely due to the fact that the phase imprinting method used to imprint solitons can also create other excitations that appear as density modulations or fringes in the BEC. Examples of such modulation can be seen in the off-diagonal images in figure~\ref{fig:soliton-sample}(c). Additional discussion of the misclassified and mislabeled data can be found in appendix~\ref{ssec:misclass}.

Our dataset includes the full-frame raw images, the cropped and rotated images as used in this study, as well as the set of the fitted integrated 2D Thomas-Fermi distribution parameters. 
This dataset is sufficient to reproduce our results but also to test fitted alternative models with varying cropping size or image resolution~\cite{solitons-data}.

\section{Conclusion and outlook}\label{sec:conclusion}
In this manuscript, we present an automated dark soliton detection and positioning system that combines ML-based image classification with standard fitting techniques to track soliton dynamics in experimental images of BECs. 
We show that the system performs on par with more traditional approaches that rely on human input for soliton identification, creating the opportunity to study soliton dynamics in large datasets. 
We also make available the first dataset of images from a dark soliton BEC experiment, which provides an opportunity for the data science community to develop more sophisticated analysis tools and to further understand nonlinear many-body physics. 

The performance of the classifier, as measured by the weighted F1 score, leaves room for improvement. 
While tuning the hyperparameters allowed us to substantially improve the initial performance, additional data is necessary to push the limits. 
However, human labeling is not only time-consuming but, as the analysis of the misclassified images revealed, is also not always reliable.
Other approaches, such as active learning ML~\cite{settles2009active}, may be more suitable for this task.
Such enlarged dataset, in turn, will enable refining the soliton classifier and perform model uncertainty quantification~\cite{abdar2020review,thiebes2020trustworthy}, which currently is not accounted for.
Together, these refinements may enable reliable in-situ deployment.

This study was preconditioned on the assumption of specific structure in the images, leading to our three classes.
Enlarged dataset will enable employing unsupervised learning strategies~\cite{ji2019invariant} to possibly discover additional classes consistent with the data without presumptions. 
This unsupervised learning of soliton-data is a prototype for ML based discovery with cold-atom data in general.

\section*{Acknowledgment}\label{sec:Acknowledgment}
This work was partially supported by NIST and NSF through the Physics Frontier Center at the JQI.
We appreciate conversations with Yuchen Yue and Justin Elenewski.

\renewcommand{\thesection}{A}
\renewcommand{\thefigure}{A\arabic{figure}}
\setcounter{figure}{0}
\renewcommand{\thetable}{A\arabic{table}}
\setcounter{table}{0}
\renewcommand{\arraystretch}{1}
\section{Appendix}\label{sec:appendix}
\subsection{Additional visualization of intermediate convolutional layers}\label{ssec:02visual}
\begin{figure}
    \flushright
    \includegraphics[width=0.9\linewidth]{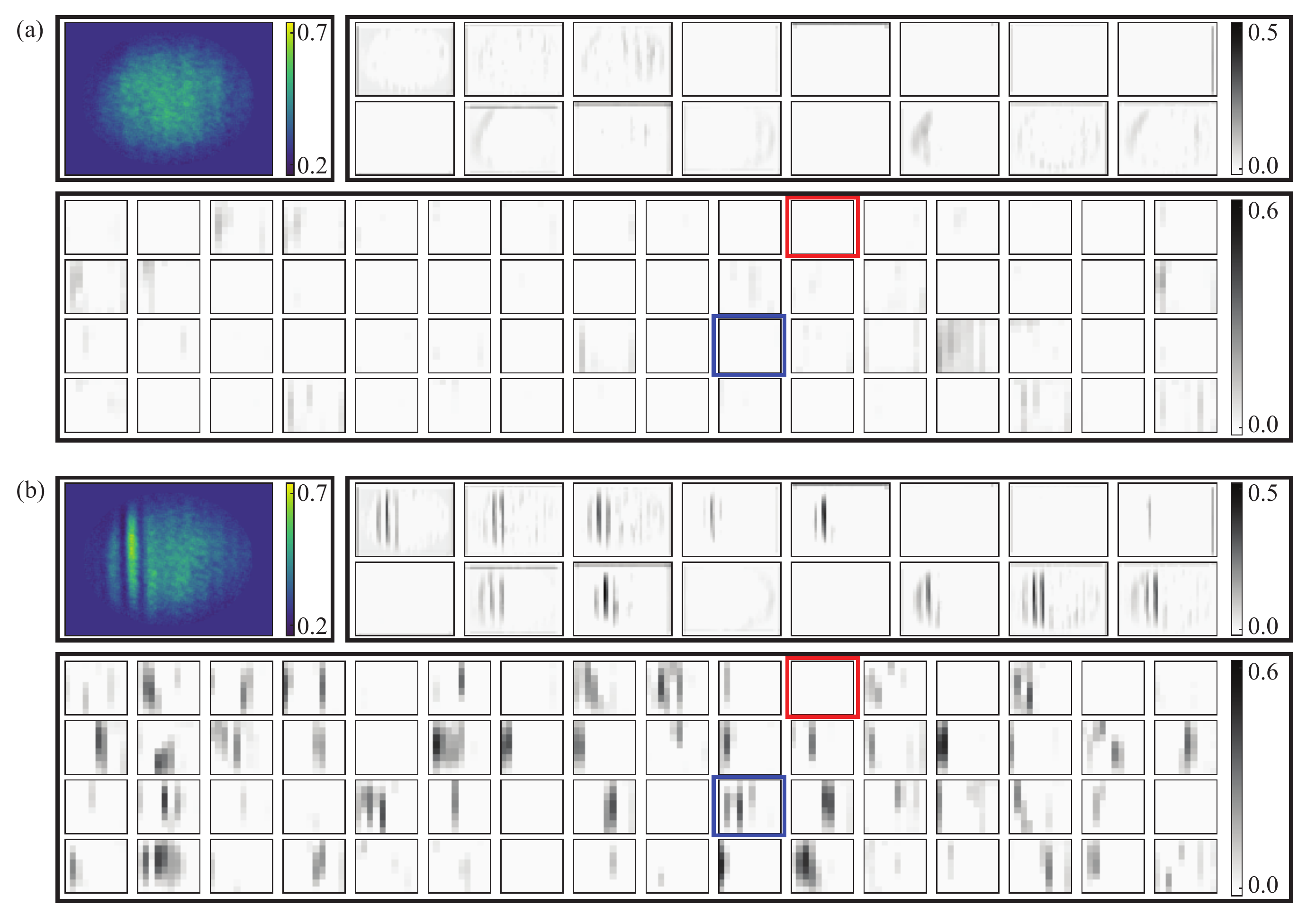}
    \caption{Visualization of the input, second, and fourth max pooling layer activation for a successfully classified (a) no soliton image and (b) other excitations one. In both cases, the top left panels are the input images, the 16 images in the top right panels represent the output of the max pooling layer, and 64 images in the bottom panels are the output of the fourth max pooling layer. The red box indicates one of the filters that captures the lone soliton feature (see figure~\ref{fig:soliton-sample}(b)). The blue boxed filter activates if more than one soliton is present.}
    \label{fig:02visual}
\end{figure}

In addition to figure~\ref{fig:soliton-sample}(b) in the main text (visualizing the single soliton case), figure~\ref{fig:02visual} shows the same intermediate layers for the correctly classified sample images from the (a) no soliton and (b) other excitations classes. 
In both figures, we highlight two filters: red box indicates a filter that activates for an image with a single soliton, while the blue box indicates a filter that activates for an image from the other excitations class. 
For an image from the no soliton class (figure~\ref{fig:02visual}(a)), neither highlighted filter is activated. 
This confirms that filters in our model are trained to properly detect and locate features that are characteristic of each class in a previously unknown image.

\subsection{Model parameter tuning}\label{ssec:tuning}
\begin{table}[!htp]\centering
\renewcommand{\arraystretch}{1.1}
\setlength{\tabcolsep}{9pt}
\scriptsize
\caption{Model parameter tuning history. For each model, we provide the number of filters for all convolutional layers (Filters), the number of nodes in the fully connected layers (Dense), as well as the kernel size of all convolutional layers ($K$), the dropout rate ($D$), the batch size ($B$), and the optimizer used of training (SGD: Stochastic gradient descent, SGD+M: SGD with moment, SGD+M+D: SGD+M with decay). 
The mean performance is averaged by 5-fold cross-validation on the training set. 
F1 score is weighted by three classes. 
Best model is highlighted. Parameters changed at each stage are highlighted in italic.}
\begin{tabular*}{\textwidth}{llrrrrrrrr}\toprule
Filters & Dense & $K$ & $D$ & $B$ & Optimizer & Weighted & Accuracy & Binary \\
 &  &  &  &  &  & F1 [\%] & [\%] & F1 [\%]\\
\midrule
8$\times$8$\times$8 &256$\times$128 &5 &0.5 &32 &Adam &65(23) &74(15) &71(18) \\
8$\times$\textit{16}$\times$\textit{32} &256$\times$128 &5 &0.5 &32 &Adam &56(23) &61(25) &63(19) \\
8$\times$16$\times$32$\times$\textit{64} &256$\times$128 &5 &0.5 &32 &Adam &67(24) &75(16) &72(20) \\
8$\times$16$\times$32$\times$64$\times$\textit{128} &256$\times$128 &5 &0.5 &32 &Adam &78(20) &82(14) &81(16) \\
8$\times$16$\times$32$\times$64$\times$128 &256$\times$128$\times$\textit{64} &5 &0.5 &32 &Adam &48(20) &62(13) &56(16) \\
8$\times$16$\times$32$\times$64$\times$128 &256$\times$64 &5 &0.5 &32 &Adam &58(25) &69(17) &65(20) \\
8$\times$16$\times$32$\times$64$\times$128 &256$\times$64$\times$\textit{16} &5 &0.5 &32 &Adam &61(19) &75(16) &69(17) \\
8$\times$16$\times$32$\times$64$\times$128 &\textit{512}$\times$\textit{128} &5 &0.5 &32 &Adam &47(20) &54(22) &56(16) \\
8$\times$16$\times$32$\times$64$\times$128 &512$\times$128$\times$\textit{32} &5 &0.5 &32 &Adam &67(24) &76(16) &72(20) \\
8$\times$16$\times$32$\times$64$\times$128 &256$\times$128$\times$64 &5 &0.5 &32 &\textit{Adamax} &86(6) &89(1) &88(3) \\
8$\times$16$\times$32$\times$64$\times$128 &256$\times$128$\times$64 &5 &0.5 &32 &\textit{SGD} &70(6) &87(2) &76(4) \\
8$\times$16$\times$32$\times$64$\times$128 &256$\times$128$\times$64 &5 &0.5 &32 &\textit{SGD+M} &64(22) &74(15) &70(18) \\
8$\times$16$\times$32$\times$64$\times$128 &256$\times$128$\times$64 &5 &0.5 &32 &\textit{SGD+M+D} &39(4) &60(9) &49(2) \\
8$\times$16$\times$32$\times$64$\times$128 &256$\times$128$\times$64 &5 &\textit{0.6} &32 &Adamax &77(5) &90(0) &85(3) \\
8$\times$16$\times$32$\times$64$\times$128 &256$\times$128$\times$64 &5 &\textit{0.7} &32 &Adamax &51(17) &68(15) &61(16) \\
8$\times$16$\times$32$\times$64$\times$128 &256$\times$128$\times$64 &5 &\textit{0.8} &32 &Adamax &44(13) &62(13) &54(12) \\
8$\times$16$\times$32$\times$64$\times$128 &256$\times$128$\times$64 &\textit{3} &0.5 &32 &Adamax &86(1) &90(1) &88(1) \\
8$\times$16$\times$32$\times$64$\times$128 &256$\times$128$\times$64 &\textit{7} &0.5 &32 &Adamax &88(1) &89(1) &90(1) \\
\rowcolor{gray}
8$\times$16$\times$32$\times$64$\times$128 &256$\times$128$\times$64 &\textit{9} &0.5 &32 &Adamax &89(0) &89(0) &90(0) \\
8$\times$16$\times$32$\times$64$\times$128 &256$\times$128$\times$64 &\textit{11} &0.5 &32 &Adamax &78(20) &82(14) &82(17) \\
8$\times$16$\times$32$\times$64$\times$128 &256$\times$128$\times$64 &9 &0.5 &\textit{16} &Adamax &79(21) &83(13) &82(17) \\
8$\times$16$\times$32$\times$64$\times$128 &256$\times$128$\times$64 &9 &0.5 &\textit{8} &Adamax &78(20) &82(13) &81(17) \\
\bottomrule
\end{tabular*}
\label{tab:3}
\end{table}

We used a naive semi-structure search approach to optimize our model parameter.
During parameter tuning, we used $k$-fold cross-validation to assess the generalizability of trained models.
For each set of hyperparameters defining the overall model (e.g., kernel size and number of layers, both convolutional and hidden), the training set was split into $k=5$ smaller sets (`folds'), four of which were used for training and one for validation. 
Once the model was trained using all five cross-validation combinations, the mean score was recorded and compared against scores from networks with other hyperparameters.
We arrange the parameters by their importance: filter number of each convolutional layers, dense layer node number, optimizer, convolution kernel size, dropout rate and batch size. 
The parameters are optimized in this order and the history of hyperparameter tuning is shown in table~\ref{tab:3}.

\subsection{Misclassified data}\label{ssec:misclass}
As discussed in section~\ref{ssec:exp}, in BEC experiments trapped condensates are often held for a certain period of time after phase imprinting. 
This is necessary to smooth out the various other excitations resulting from the phase imprinting process. 
However, by looking only at a single image (single holding time), all the information about the soliton dynamics is lost, and other excitations can be confused with shallow solitons.
In the final 640 images in the test dataset, there are 68 misclassified images in one of six possible ways.
As can be seen in figure~\ref{fig:soliton-sample}(c), for our model, the most confusion comes from distinguishing between the single soliton and other excitations classes. 
Upon reviewing the 58 images misclassified in this way, we find that out of 27 images classified as other excitations only two clearly contain lone solitons. 
The remaining 25 images are confusing and thus should belong to the other excitations class. 
Interestingly, for this case, the classifier seems to be nearly as likely to be wrong as confused (see middle columns in figure ~\ref{fig:class-result}(b)).
All 31 images classified as single solitons are true misclassification. For these images, the classifier is  confidently wrong (see last columns in figure ~\ref{fig:class-result}(b)).
We suspect that it might be possible to improve the classifier by training with less penalty for classifying an ambiguous image to other excitations class.
This also suggests that active learning strategy might be better for training model and labeling data than relaying on dataset labeled by humans.

\bibliographystyle{unsrt}

\begin{thebibliography}{10}

\bibitem{Aurisano_2016}
A.~Aurisano, A.~Radovic, D.~Rocco, A.~Himmel, M.D. Messier, E.~Niner,
  G.~Pawloski, F.~Psihas, A.~Sousa, and P.~Vahle.
\newblock A convolutional neural network neutrino event classifier.
\newblock {\em Journal of Instrumentation}, 11(09):P09001--P09001, sep 2016.

\bibitem{Acciarri_2017}
R.~Acciarri, C.~Adams, R.~An, J.~Asaadi, and M.~Auger {\it et al}.
\newblock Convolutional neural networks applied to neutrino events in a liquid
  argon time projection chamber.
\newblock {\em Journal of Instrumentation}, 12(03):P03011--P03011, mar 2017.

\bibitem{kagan2020imagebased}
Michael Kagan.
\newblock Image-based jet analysis.
\newblock {\em arXiv preprint arXiv:2012.09719}, 2020.

\bibitem{Golovatiuk_2020}
Artem Golovatiuk, Giovanni~De Lellis, and Andrey Ustyuzhanin.
\newblock Deep learning for directional dark matter search.
\newblock {\em Journal of Physics: Conference Series}, 1525:012108, apr 2020.

\bibitem{Khosa_2020}
Charanjit~K Khosa, Lucy Mars, Joel Richards, and Veronica Sanz.
\newblock Convolutional neural networks for direct detection of dark matter.
\newblock {\em Journal of Physics G: Nuclear and Particle Physics},
  47(9):095201, jul 2020.

\bibitem{Kalantre17-MLD}
S.~S. Kalantre, Justyna~P. Zwolak, Stephen Ragole, Xingyao Wu, Neil~M.
  Zimmerman, M.~D. Stewart, and Jacob~M. Taylor.
\newblock Machine learning techniques for state recognition and auto-tuning in
  quantum dots.
\newblock {\em npj Quantum Information}, 5(6):1--10, 2017.

\bibitem{Mills19-CAT}
A.~R. Mills, M.~M. Feldman, C.~Monical, P.~J. Lewis, K.~W. Larson, A.~M.
  Mounce, and J.~R. Petta.
\newblock Computer-automated tuning procedures for semiconductor quantum dot
  arrays.
\newblock {\em Applied Physics Letters}, 115(11):113501, 2019.

\bibitem{Zwolak20-AQD}
Justyna~P. Zwolak, Thomas McJunkin, Sandesh~S. Kalantre, J.P. Dodson, E.R.
  MacQuarrie, D.E. Savage, M.G. Lagally, S.N. Coppersmith, Mark~A. Eriksson,
  and Jacob~M. Taylor.
\newblock Autotuning of double-dot devices in situ with machine learning.
\newblock {\em Phys. Rev. Applied}, 13:034075, Mar 2020.

\bibitem{Usman2020}
Muhammad Usman, Yi~Zheng Wong, Charles~D. Hill, and Lloyd C.~L. Hollenberg.
\newblock Framework for atomic-level characterisation of quantum computer
  arrays by machine learning.
\newblock {\em npj Computational Materials}, 6(1):1--8, 2020.

\bibitem{cao2019convolutional}
Zhuo Cao, Yabo Dan, Zheng Xiong, Chengcheng Niu, Xiang Li, Songrong Qian, and
  Jianjun Hu.
\newblock Convolutional neural networks for crystal material property
  prediction using hybrid orbital-field matrix and magpie descriptors.
\newblock {\em Crystals}, 9(4):191, 2019.

\bibitem{PhysRevMaterials.4.093801}
Mohammadreza Karamad, Rishikesh Magar, Yuting Shi, Samira Siahrostami, Ian~D.
  Gates, and Amir Barati~Farimani.
\newblock Orbital graph convolutional neural network for material property
  prediction.
\newblock {\em Phys. Rev. Materials}, 4:093801, Sep 2020.

\bibitem{gubernatis2018machine}
JE~Gubernatis and T~Lookman.
\newblock Machine learning in materials design and discovery: Examples from the
  present and suggestions for the future.
\newblock {\em Physical Review Materials}, 2(12):120301, 2018.

\bibitem{duvenaud2015convolutional}
David~K Duvenaud, Dougal Maclaurin, Jorge Iparraguirre, Rafael Bombarell,
  Timothy Hirzel, Al{\'a}n Aspuru-Guzik, and Ryan~P Adams.
\newblock Convolutional networks on graphs for learning molecular fingerprints.
\newblock {\em Advances in neural information processing systems},
  28:2224--2232, 2015.

\bibitem{butler2018machine}
Keith~T Butler, Daniel~W Davies, Hugh Cartwright, Olexandr Isayev, and Aron
  Walsh.
\newblock Machine learning for molecular and materials science.
\newblock {\em Nature}, 559(7715):547--555, 2018.

\bibitem{winter2019learning}
Robin Winter, Floriane Montanari, Frank No{\'e}, and Djork-Arn{\'e} Clevert.
\newblock Learning continuous and data-driven molecular descriptors by
  translating equivalent chemical representations.
\newblock {\em Chemical Science}, 10(6):1692--1701, 2019.

\bibitem{Rem19-PTN}
Benno~S. Rem, Niklas Käming, Matthias Tarnowski, Luca Asteria, Nick
  Fläschner, Christoph Becker, Klaus Sengstock, and Christof Weitenberg.
\newblock Identifying quantum phase transitions using artificial neural
  networks on experimental data.
\newblock {\em Nature Physics}, page~1, July 2019.

\bibitem{PhysRevApplied.14.014011}
Gal Ness, Anastasiya Vainbaum, Constantine Shkedrov, Yanay Florshaim, and Yoav
  Sagi.
\newblock Single-exposure absorption imaging of ultracold atoms using deep
  learning.
\newblock {\em Phys. Rev. Applied}, 14:014011, Jul 2020.

\bibitem{Pilati19-SMU}
S.~Pilati and P.~Pieri.
\newblock Supervised machine learning of ultracold atoms with speckle disorder.
\newblock {\em Scientific Reports}, 9:5613, Jul 2019.

\bibitem{metz2021deep}
Friederike Metz, Juan Polo, Natalya Weber, and Thomas Busch.
\newblock Deep learning based quantum vortex detection in atomic bose-einstein
  condensates.
\newblock {\em Machine Learning: Science and Technology}, 2021.

\bibitem{PhysRevLett.101.130401}
A.~Weller, J.~P. Ronzheimer, C.~Gross, J.~Esteve, M.~K. Oberthaler, D.~J.
  Frantzeskakis, G.~Theocharis, and P.~G. Kevrekidis.
\newblock Experimental observation of oscillating and interacting matter wave
  dark solitons.
\newblock {\em Phys. Rev. Lett.}, 101:130401, Sep 2008.

\bibitem{Frantzeskakis_2010}
D~J Frantzeskakis.
\newblock Dark solitons in atomic {Bose}{\textendash}{Einstein} condensates:
  from theory to experiments.
\newblock {\em Journal of Physics A: Mathematical and Theoretical},
  43(21):213001, may 2010.

\bibitem{Russel1837}
J.~S. Russel.
\newblock {\em Report of the Committee on Waves}, pages 417--468.
\newblock British reports VI, 1837.
\newblock plus plates 1–8.

\bibitem{Osborne1980}
A.~R. Osborne and T.~L. Burch.
\newblock Internal solitons in the andaman sea.
\newblock {\em Science}, 208(4443):451--460, 1980.

\bibitem{Lakshmanan2009}
M.~Lakshmanan.
\newblock {\em Solitons, Tsunamis and Oceanographical Applications of}, pages
  8506--8521.
\newblock Springer New York, New York, NY, 2009.

\bibitem{Burger1999}
S.~Burger, K.~Bongs, S.~Dettmer, W.~Ertmer, K.~Sengstock, A.~Sanpera, G.~V.
  Shlyapnikov, and M.~Lewenstein.
\newblock Dark solitons in {Bose}-{Einstein} condensates.
\newblock {\em Phys. Rev. Lett.}, 83:5198--5201, Dec 1999.

\bibitem{Denschlag97}
J.~Denschlag, J.~E. Simsarian, D.~L. Feder, Charles~W. Clark, L.~A. Collins,
  J.~Cubizolles, L.~Deng, E.~W. Hagley, K.~Helmerson, W.~P. Reinhardt, S.~L.
  Rolston, B.~I. Schneider, and W.~D. Phillips.
\newblock Generating solitons by phase engineering of a {Bose}-{Einstein}
  condensate.
\newblock {\em Science}, 287(5450):97--101, 2000.

\bibitem{Hasegawa1973}
Akira Hasegawa and Frederick Tappert.
\newblock Transmission of stationary nonlinear optical pulses in dispersive
  dielectric fibers. ii. normal dispersion.
\newblock {\em Applied Physics Letters}, 23(4):171--172, 1973.

\bibitem{PhysRevLett.45.1095}
L.~F. Mollenauer, R.~H. Stolen, and J.~P. Gordon.
\newblock Experimental observation of picosecond pulse narrowing and solitons
  in optical fibers.
\newblock {\em Phys. Rev. Lett.}, 45:1095--1098, Sep 1980.

\bibitem{Stasiewicz2003}
K.~Stasiewicz, P.~K. Shukla, G.~Gustafsson, S.~Buchert, B.~Lavraud, B.~Thid\'e,
  and Z.~Klos.
\newblock Slow magnetosonic solitons detected by the cluster spacecraft.
\newblock {\em Phys. Rev. Lett.}, 90:085002, Feb 2003.

\bibitem{Hashizume1985}
Yasuo Hashizume.
\newblock Nonlinear pressure waves in a fluid-filled elastic tube.
\newblock {\em Journal of the Physical Society of Japan}, 54(9):3305--3312,
  1985.

\bibitem{Yomosa1984}
Sigeo Yomosa.
\newblock Solitary waves in large blood vessels.
\newblock {\em Journal of the Physical Society of Japan}, 56(2):506--520, 1987.

\bibitem{MOLLENAUER20061}
Linn~F. Mollenauer and James~P. Gordon.
\newblock {\em Solitons in Optical Fibers}.
\newblock Academic Press, Burlington, 2006.

\bibitem{902164}
A.~{Hasegawa}.
\newblock Soliton-based optical communications: an overview.
\newblock {\em IEEE Journal of Selected Topics in Quantum Electronics},
  6(6):1161--1172, 2000.

\bibitem{Aycock2503}
Lauren~M. Aycock, Hilary~M. Hurst, Dmitry~K. Efimkin, Dina Genkina, Hsin-I Lu,
  Victor~M. Galitski, and I.~B. Spielman.
\newblock Brownian motion of solitons in a {Bose}{\textendash}{Einstein}
  condensate.
\newblock {\em Proceedings of the National Academy of Sciences},
  114(10):2503--2508, 2017.

\bibitem{Fritsch20-SCV}
A.~R. Fritsch, Mingwu Lu, G.~H. Reid, A.~M. Pi\~neiro, and I.~B. Spielman.
\newblock Creating solitons with controllable and near-zero velocity in
  {Bose}-{Einstein} condensates.
\newblock {\em Phys. Rev. A}, 101:053629, May 2020.

\bibitem{rawat2017deep}
Waseem Rawat and Zenghui Wang.
\newblock Deep convolutional neural networks for image classification: A
  comprehensive review.
\newblock {\em Neural computation}, 29(9):2352--2449, 2017.

\bibitem{solitons-data}
Database: data.nist.gov [Internet].
\newblock Dark solitons in {BEC}s dataset, 2020.
\newblock Available from: \url{https://doi.org/10.18434/mds2-2363}.

\bibitem{Lin2009}
Y.-J. Lin, A.~R. Perry, R.~L. Compton, I.~B. Spielman, and J.~V. Porto.
\newblock Rapid production of [sup 87]rb bose-einstein condensates in a
  combined magnetic and optical potential.
\newblock {\em Physical Review A (Atomic, Molecular, and Optical Physics)},
  79(6):63631, 2009.

\bibitem{PhysRevLett.89.110401}
A.~Muryshev, G.~V. Shlyapnikov, W.~Ertmer, K.~Sengstock, and M.~Lewenstein.
\newblock Dynamics of dark solitons in elongated {Bose}-{Einstein} condensates.
\newblock {\em Phys. Rev. Lett.}, 89:110401, Aug 2002.

\bibitem{Mateo2015}
A~Mu{\~n}oz Mateo and J~Brand.
\newblock Stability and dispersion relations of three-dimensional solitary
  waves in trapped bose--einstein condensates.
\newblock {\em New Journal of Physics}, 17(12):125013, 2015.

\bibitem{ketterle1999making}
W.~Ketterle, D.~S. Durfee, and D.~M. Stamper-Kurn.
\newblock Making, probing and understanding {Bose}-{Einstein} condensates.
\newblock In M.~Inguscio, S.~Stringari, and C.~E. Wieman, editors, {\em
  Bose-Einstein condensation in atomic gases, Proceedings of the International
  School of Physics ``Enrico Fermi", Course CXL}, pages 67--176. IOS Press,
  Amsterdam, 1999.

\bibitem{Castin1996}
Y.~Castin and R.~Dum.
\newblock Bose-einstein condensates in time dependent traps.
\newblock {\em Phys. Rev. Lett.}, 77(27):5315--5319, December 1996.

\bibitem{scikit-learn}
F.~Pedregosa, G.~Varoquaux, A.~Gramfort, V.~Michel, B.~Thirion, O.~Grisel,
  M.~Blondel, P.~Prettenhofer, R.~Weiss, V.~Dubourg, J.~Vanderplas, A.~Passos,
  D.~Cournapeau, M.~Brucher, M.~Perrot, and E.~Duchesnay.
\newblock Scikit-learn: Machine learning in {P}ython.
\newblock {\em Journal of Machine Learning Research}, 12:2825--2830, 2011.

\bibitem{settles2009active}
Burr Settles.
\newblock Active learning literature survey.
\newblock Technical report, University of Wisconsin-Madison Department of
  Computer Sciences, 2009.

\bibitem{abdar2020review}
Moloud Abdar, Farhad Pourpanah, Sadiq Hussain, Dana Rezazadegan, Li~Liu,
  Mohammad Ghavamzadeh, Paul Fieguth, Abbas Khosravi, U~Rajendra Acharya,
  Vladimir Makarenkov, et~al.
\newblock A review of uncertainty quantification in deep learning: Techniques,
  applications and challenges.
\newblock {\em arXiv preprint arXiv:2011.06225}, 2020.

\bibitem{thiebes2020trustworthy}
Scott Thiebes, Sebastian Lins, and Ali Sunyaev.
\newblock Trustworthy artificial intelligence.
\newblock {\em Electronic Markets}, pages 1--18, 2020.

\bibitem{ji2019invariant}
Xu~Ji, Jo{\~a}o~F Henriques, and Andrea Vedaldi.
\newblock Invariant information clustering for unsupervised image
  classification and segmentation.
\newblock In {\em Proceedings of the IEEE International Conference on Computer
  Vision}, pages 9865--9874, 2019.

\end{thebibliography}
\newcommand{\newblock}{}

\end{document}